# Broadband terahertz pulse emission from ZnGeP$_2$

J. D. Rowley,[1] J. K. Pierce,[1] A. T. Brant,[1] L. E. Halliburton,[1] N. C. Giles,[2] P. G. Schunemann,[3] A. D. Bristow,[1,*]

[1]*Department of Physics, West Virginia University, Morgantown, WV, 26506-6315, USA*
[2]*Department of Engineering Physics, Air Force Institute of Technology, Wright-Patterson Air Force Base, Ohio 45433, USA*
[3]*BAE Systems, MER15-1813, P.O. Box 868, Nashua, New Hampshire 03061, USA*
*\*Corresponding author: alan.bristow@mail.wvu.edu*

Optical rectification is demonstrated in (110)-cut ZnGeP$_2$ (ZGP) providing broadband terahertz (THz) generation. The source is compared to both GaP and GaAs over a wavelength range of 1150 to 1600 nm and peak intensity range of 0.5 GW/cm$^2$ to 40 GW/cm$^2$. ZGP peak-to-peak field amplitude is larger than in the other materials due to either lower nonlinear absorption or larger second order nonlinearity. This material is well suited for broadband THz generation across a wide range of infrared excitation wavelengths.

Over the last decade, broadband sources of intense THz pulses have received significant attention due to the possibility of accessing a diverse range of light-matter interactions and the potential for creating high-speed electronics. For a recent review of the progress see Ref. [1] and references therein. Large-area zincblende $\bar{4}3m$ crystals are commonly used in time-domain terahertz spectroscopy as sources and electro-optic (EO) samplers [2] due to large second-order nonlinear coefficients. A common issue with materials like GaAs and ZnTe is that the optimum wavelength for THz and optical velocity matching is poorly positioned with respect to the onset of two-photon absorption.

Here we generate broadband THz pulses in ZnGeP$_2$, a chalcopyrite (pnictide) crystal with $\bar{4}2m$ symmetry, by optical rectification. Optical rectification is a second-order nonlinear optical process that generates a transient D.C. polarization, which typically creates THz radiation for short excitation pulses. In previous work, ZnGeP$_2$ has only been used as a source of narrow-band and tuneable THz radiation through parametric down conversion [3,4]. This work verifies that ZGP has a suitable coherence length, high laser-damage threshold [3] and nonlinear figure-of-merit [5] for optical rectification. ZGP's generation efficiency surpasses that of high-quality and high-resistivity wafers of GaAs and GaP across the infrared, including the datacoms C-band around 1550 nm. This study complements the recent resurgence of interest in ZGP as a nonlinear material driven by high-quality growth [6,7].

Undoped ZGP single crystals were grown by the horizontal gradient freeze technique [6]. Samples were cut in the (110) plane and double-side polished for optical transmission measurements. Sample thicknesses are 0.33 mm and 0.93 mm for the short and long crystals respectively. The orientation of the crystals was verified by electron paramagnetic resonance (EPR), ensuring that the z-axis is in the plane of the plate. EPR measures the phosphorus hyperfine signal associated with the zinc-vacancy acceptor in as-grown ZGP, differentiating the ⟨100⟩ and ⟨001⟩ directions [8]. ZGP is birefringent due to a ~2% lattice compression in the z-axis, which complicates the THz generation for pump polarizations that are not parallel or perpendicular to this axis. Measurements were performed at normal incidence with the crystal rotated in the plane to maximize the THz signal. The rotation results in the pump being polarized perpendicular to the z-axis of the ZGP crystal, such that the electric field is ordinary (o-wave), the crystal does not act as a waveplate, and the generated THz is extraordinary (e-wave). A full study of the polarization-dependent THz generation is not shown in this letter. Non-birefringent reference crystals of undoped, high resistivity [110]-cut GaP (0.13 mm) and GaAs (0.5 mm) are also measured.

Figure 1(a) shows the time-domain THz-emission spectroscopy setup [9], which uses 100-fs pulses from a 1 KHz regenerative laser amplifier. "Signal" pulses from an optical parametric amplifier (OPA) impinge the test nonlinear crystals (NLC), generating THz radiation that is detected by EO sampling. The gate pulse is centered at 800 nm, and a detection sensitivity of ~1×10$^{-8}$ is achieved using a Soleil-Babinet compensator (SBC), Wollaston prism (WP) and balanced (A-B) biased-photodiodes operating close to the shot-noise limit. The excitation pulses are

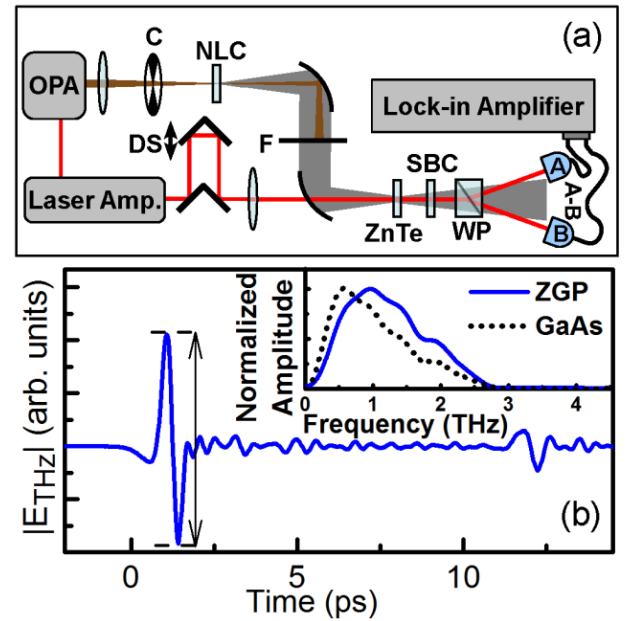

Fig. 1 (color online) (a) Time-domain terahertz (THz) emission spectroscopy setup. (b) Typical THz transient for ZnGeP$_2$ (ZGP) pumped at 1300 nm. Inset shows normalized power spectra for ZGP and GaAs.

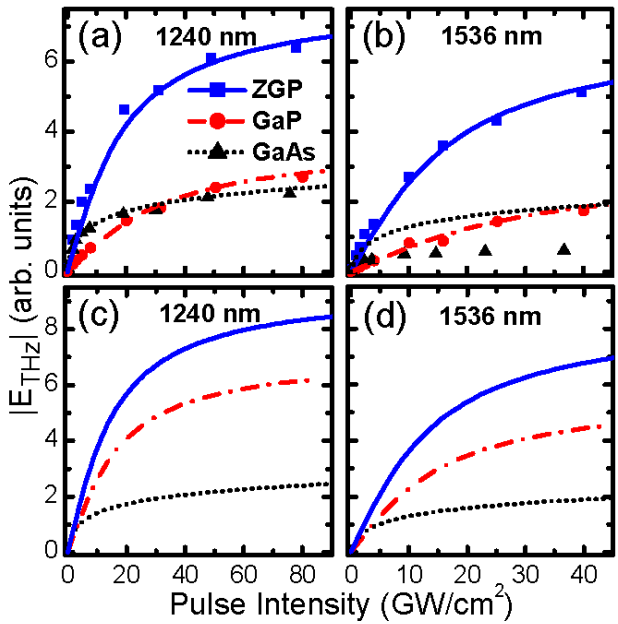

Fig. 2 (color online) Intensity-dependent peak-to-peak THz amplitude for ZGP (0.33 mm), GaP (0.13 mm) and GaAs (0.5 mm) at (a) 1240 nm and (b) 1536 nm. Estimated THz amplitude adjusting for the crystal length, based on a thickness of 0.5 mm for (c) 1240 nm and (d) 1536 nm.

mechanically chopped (C) for phase-sensitive detection and are tunable from 1150 nm to 1600 nm. A long-pass polyvinyl chloride filter (F) prevents the pump light from reaching the EO sampler (ZnTe, 0.5-mm thick).

Figure 1(b) shows a typical THz transient mapped out by the gate pulse when scanning the relative time delay (delay stage, DS). The EO signal is linearly proportional to the THz field $|E_{THz}|$ and the signal yields >100 signal-to-noise ratio. The data reveals a main THz transient and an echo at 12 ps. Additional oscillations are due to the the 3-THz bandwidth of the ZnTe sampling crystal and water absorption, even with a nitrogen purge of the setup. Normalized power spectra were obtained by a fast Fourier transform with a 3-ps window function in analysis; see the inset of the figure. As expected, the spectrum is broadband (0.1 THz to 3 THz). The THz emission from ZGP is not limited by absorption due to IR-active phonons. Absorption in this frequency range is known to be smaller than that of ZnTe and similar to that of GaAs [5]. By comparison the spectrum of GaAs peaks at lower frequencies, suggesting that the peak-to-peak electric field strength will be smaller than that obtained from ZGP [10].

Figure 2 shows the pump peak-intensity dependence of the THz amplitude $|E_{THz}|$ for the 0.33-mm ZGP, 0.13-mm GaP and 0.5-mm GaAs samples, pumped at (a) 1260 nm and (b) 1536 nm. $|E_{THz}|$ is measured as the peak-to-peak voltage change during the main transient, as indicated in Fig. 1(b). Peak-intensity values are corrected to account for reflection losses at the input interface. At low intensities, $|E_{THz}|$ increases linearly because $dE_{THz}/dz \propto d_{36}I(z)$, where $d_{36}$ is the nonlinear tensor element. The effective coefficient $d_{eff}$ and $d_{36}$ are interchangeable in the proportionality because all crystals are identically cut. This proportionality is applicable for birefringent ZGP because the pump electric field is purely o-wave. At higher intensities, the signal saturates due to nonlinear absorption of the excitation pulse, which follows a generalized form of Beer's law: $dI(z)/dz = -\alpha I(z) - \beta I^2(z) - \gamma I^3(z) + \cdots$, where $\alpha$ is the linear absorption coefficient that is measured in an auxiliary measurement, and $\beta, \gamma$ are the two- and three-photon (2PA and 3PA) absorption coefficients. Values used to fit the experimental data are shown in Table I. For GaAs, the absorption is dominated by 2PA, whereas for the other crystals 3PA dominates. Even neglecting free-carrier absorption of the THz and deviations from perfect phase matching, the fitting parameters are in good agreement with values in the literature [11-14], with the exception of the GaAs response at 1536 nm. For this data, poor phase-matching [15] lowers the observed response compared to the simulation.

TABLE I. Optical properties of the various semiconductors crystal at 1240(1536) nm, where × means the term is neglected in fitting the intensity dependence.

|  | $d_{36}$ (pm/V) | $\alpha$ (cm$^{-1}$) | $\beta$ (cm/GW) | $\gamma$ (cm$^3$/GW$^2$) |
|---|---|---|---|---|
| GaAs | 130 (119) | 1 | 35 | × |
| GaP | 50 (45) | 3.5 | × | 0.08 (0.10) |
| ZGP | 70 | 1.8 | × | 0.09 (0.11) |

The difference in crystal length for the relative peak-to-peak $|E_{THz}|$ presented in Fig. 2(a) and (b) skews the results in favor of the thicker crystal. Consequently, an estimate of the respective curve shapes and strengths are given in Fig. 2(c) and (d) for the same excitation conditions, but where each crystal is taken to be 0.5 mm thick. After compensating for crystal length, the ZGP still out performs both GaAs and GaP at the wavelengths shown. The trend of ZGP's higher performance is also observed at 1342 nm and 1442 nm (data not shown).

Figure 3 (a) shows the photon energy (wavelength) dependence of the relative peak-to-peak $|E_{THz}|$ from ZGP of two thicknesses, GaP and GaAs. The above intensity dependence is used to normalize the data at each wavelength, accounting for variations in the OPA emission and the transmission losses through the optical excitation setup. Peak pump intensity of ~8 GW/cm$^2$ was used to minimize the effects of nonlinear absorption. Consequently, variations in $|E_{THz}|$ due to differences in 3PA can be neglected, but effects of crystal thickness, velocity matching and 2PA cannot. The short ZGP and GaP samples are mostly unaffected by photon energy, showing a flat response. The long ZGP and GaAs data however, decrease at lower photon energies and the ZGP response peaks at 1200 nm, due to a combination of velocity mismatch in the THz and pump pulses and nonlinear absorption of the pump.

The velocity matching is quantified by the coherence length $l_c = c/(2\nu_{THz}|n_{THz} - n_g|)$, where $c$ is the speed of light, $\nu_{THz}$ and $n_{THz}$ are the phase frequency and refractive index of the THz, and $n_g = n - \lambda \partial n/\partial \lambda$ is the optical group index for the pump pulse with central optical wavelength $\lambda$ and corresponding linear refractive index $n$ [15]. Figure 3(b) shows the coherence length for ZGP, calculated using a published dispersion relation for the o-wave pump [16] and a single value ($n_e$ = 3.41) for the THz refractive index [17] at ~1.5 THz. Figure 3(b) also shows the normalized wavelength dependence of $\beta$ and $\gamma$ for ZGP [18]. Comparison of the $l_c$ and $\beta$ to the experimental data for the long ZGP crystal suggest that the long and short wavelength response are limited by the velocity matching and 2PA respectively, leading to the observed peak.

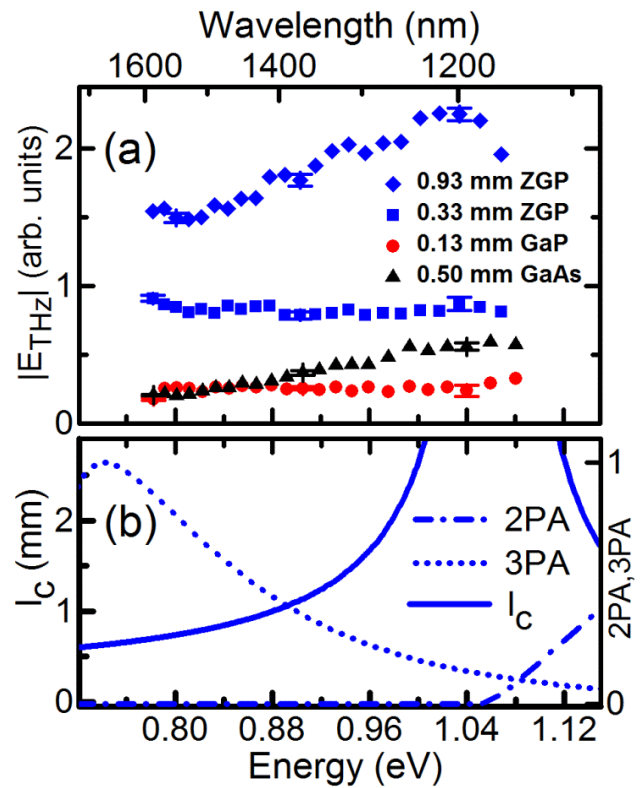

Fig. 3 (color online) (a) Photon energy and wavelength dependence of the peak-to-peak THz amplitude for ZGP, GaP and GaAs. (b) Simulation of the wavelength-dependent coherence length and *normalized* two- and three-photon absorption coefficients in ZGP.

Moderate length ZGP crystals (<1 mm) appear to be efficient THz sources across the infrared because of convenient velocity matching. Therefore, ZGP offers a better alternative than GaAs excited in the telecoms/datacoms window [15], due to lower multiphoton absorption. It may also operate more efficiently than GaP excited at its optimum coherence length [19], due to larger $d_{eff}$ and comparable $\beta$, and $\gamma$ [5].

In summary, it has been shown experimentally that ZnGeP$_2$ is capable of effectively producing broadband THz pulses by optical rectification and that crystal quality and birefringence are not impediments to the generation process. While neither the source nor the setup in this study are optimized to compare directly with large-area optical rectification, tilted wave-front, or air-plasma THz sources [10], the observed response encourages further exploration of THz generation in highly nonlinear chalcopyrite crystals by optical rectification.

JDR wishes to thank the WVNano Initiative for support. The views expressed in this article are those of the authors and do not necessarily reflect the official policy or position of the Air Force, the Department of Defense, or the United States Government.